\documentclass[
reprint,
superscriptaddress,
nofootinbib,
amsmath,amssymb,
 aps,
]{revtex4-2}

\usepackage{graphicx}
\usepackage{dcolumn}
\usepackage{bm}
\usepackage{xcolor}
\usepackage{stmaryrd}
\usepackage{times}
\usepackage{BOONDOX-calo}

\begin{document}

\preprint{APS/123-QED}

\title{Microscopic Theory of a Fluctuation-Induced Dynamical Crossover in Supercooled Liquids}

\author{Corentin C. L. Laudicina}
\affiliation{Soft Matter and Biological Physics, Department of Applied Physics, Eindhoven University of Technology,
P.O. Box 513, 5600 MB Eindhoven, Netherlands
}
\author{Liesbeth M. C. Janssen}
\affiliation{Soft Matter and Biological Physics, Department of Applied Physics, Eindhoven University of Technology,
P.O. Box 513, 5600 MB Eindhoven, Netherlands
}
\affiliation{Institute for Complex Molecular Systems, Eindhoven University of Technology, P.O. Box 513, 5600MB Eindhoven, The Netherlands}
\author{Grzegorz Szamel}
\affiliation{Department of Chemistry, Colorado State University, Fort Collins, Colorado 80523, USA}

\date{\today}
\begin{abstract}
    Mean-field theories of the glass transition predict a phase transition to a dynamically arrested state, yet no such transition is observed in experiments or simulations of finite-dimensional systems. We resolve this long-standing discrepancy by incorporating critical dynamical fluctuations into a microscopic mode-coupling framework. We show that these fluctuations round off the mean-field singularity and restore ergodicity at all finite densities (or temperatures) without invoking activated dynamics or facilitation. The resulting effective theory describes the order parameter as a stochastic process with self-induced, annealed disorder, determined self-consistently at the mean-field level. In the $\beta$-relaxation regime it reduces to stochastic beta-relaxation theory, thereby unifying mode-coupling and replica-based approaches beyond mean-field. All parameters of the stochastic beta-relaxation theory are fixed by the static structure, enabling parameter-free predictions that extend mean-field theory into finite dimensions.
\end{abstract}
\maketitle

Many materials exhibit a dramatic slowdown of relaxation upon cooling or compression: relaxation times grow by many orders of magnitude despite only subtle structural changes \cite{debenedetti2001supercooled, donth2013glass}. This disconnect between structure and dynamics lies at the heart of the longstanding `glass transition problem' \cite{anderson1995through,biroli2013perspective,dauchot2023glass}. The dramatic slowdown is accompanied by so-called dynamic heterogeneity, i.e., transient, spatially correlated regions of high and low mobility whose characteristic size and lifetime grow as vitrification is approached \cite{berthier2011overview}.

Dynamic heterogeneity is often interpreted as critical dynamical fluctuations associated with a critical point \cite{dasgupta1991there, yamamoto1998dynamics, berthier2004time, berthier2007spontaneousI, berthier2007spontaneousII}, as predicted by mean-field theories such as the static replica approach \cite{parisi2010mean,mezard2012glasses} and the dynamical mode-coupling theory (MCT) of the glass transition \cite{leutheusser1984dynamical, bengtzelius1984dynamics, gotze2009complex}. At this mean-field critical point, ergodicity is lost and the system enters a dynamically arrested state, with diverging susceptibilities and correlation lengths as in a standard phase transition \cite{franz2000onnon, donati2002theory, biroli2004diverging,biroli2006inhomogeneous}.

However, neither experiments \cite{brambilla2009probing, mallamace2010transport,schmidtke2012boiling} nor simulations \cite{berthier2020finite, scalliet2022thirty, charbonneau2022dimensional} observe a true dynamical arrest; dynamics remain ergodic, with only modest growth of susceptibilities and correlation lengths \cite{glotzer2000time, lavcevic2002growing, berthier2005direct, dauchot2005dynamical, mosayebi2010probing,FlennerSzamel2010PRL,flenner2012characterizing, karmakar2014growing, tah2020signature}. This mismatch has prompted alternative approaches that either abandon finite temperature criticality \cite{tarjus1995breakdown,olsen1998structural, garrahan2002geometrical, whitelam2004dynamic, lerner2009statistical, ozawa2023elasticity, dyre2024solid} or attribute the absence of a transition to a variety of ergodicity-restoring dynamical processes often called `activated processes' \cite{Schweizer2003NLE, mayer2006cooperativity, MedinaNoyola2007PRE, bhattacharyya2008facilitation, chong2008connections, janssen2015microscopic, Schweizer2013EC}. Still, a universally accepted microscopic theory rationalizing the absence of a transition is lacking.

Any beyond-mean-field description of phase transitions requires a self-consistent treatment of critical fluctuations of the order parameter; these typically renormalize mean-field predictions but can also qualitatively change the transition \cite{halperin1974first,brazovskii1975phase}. In the context of the glass transition problem, replica-based field-theoretical approaches showed that critical fluctuations can destabilize the mean-field glass transition altogether \cite{franz2011field,franz2012quantitative}. Yet, those analyses are essentially static, with dynamical aspects of the problem only obtainable through a mapping of replicated field theories onto dynamical supersymmetric field theories \cite{rizzo2016dynamical}. A satisfactory description demands a microscopic dynamical theory that incorporates critical fluctuations beyond mean-field.

To approach this problem, let us first recall that MCT is formulated in terms of the intermediate scattering function, $F(k,t)$, a two-point time-dependent density correlator. Its long-time limit, the Debye-Waller factor $F(k) = \lim_{t\to\infty}F(k,t)$, vanishes in the ergodic phase, is finite in the arrested one, and thus serves as the order parameter for the glass transition. Since MCT exclusively involves $F(k,t)$, it neglects spatio-temporal fluctuations of local density relaxation processes, discarding the physics giving rise to dynamic heterogeneity. These fluctuations, which become critical at the mean-field transition, are naturally encoded in four-point functions, in particular in the four-point dynamic structure factor, $S_4(\boldsymbol{q} ; \boldsymbol{k}, t)$, which measures spatial correlations  on lengthscales $2\pi/|\boldsymbol{q}|$ of fluctuations in relaxation processes probed at wavevector $\boldsymbol{k}$. Its $|\boldsymbol{q}|\to0$ limit defines the well-studied dynamic susceptibility $\chi_4(k,t)$ \cite{berthier2011overview}. 

An important advance made by Berthier \emph{et al.}~\cite{berthier2007spontaneousI, berthier2007spontaneousII} was to establish that the microscopic mechanisms driving the critical behavior of $S_4$ are encoded in the so-called three-point susceptibilities $\chi_{\boldsymbol{q}}(\boldsymbol{k}, t)$, which measure the response of the order parameter $F(k,t)$ to a spatially modulated perturbation at wavevector $\boldsymbol{q}$ \cite{biroli2006inhomogeneous}. An incorporation of critical fluctuations around the dynamical glass transition can therefore be achieved by including $\chi_{\boldsymbol{q}}(\boldsymbol{k}, t)$, calculated within mode-coupling approximation, into a theoretical analysis of the order parameter.

Using a diagrammatic formulation of MCT \cite{szamel2007dynamics}, we develop a controlled microscopic treatment of critical fluctuations in the vicinity of the dynamical glass transition, incorporating the three-point susceptibilities $\chi_{\boldsymbol{q}}(\boldsymbol{k}, t)$ mentioned above. We show that their inclusion transforms the mean-field singularity into a dynamical crossover, without invoking any additional physical process. The key steps, and main results are presented here, with complete diagrammatic rules, detailed derivations, and technical calculations deferred to a companion paper \cite{companionpaper}.

\emph{Outline of the Computation}\textemdash{}Our approach consists of two steps. First, since the critical behavior is governed by the $\beta$-regime of relaxation\footnote{Recall, glassy relaxation exhibits two successive dynamical regimes: the $\beta$-regime corresponding dynamics around the plateau of $F(k,t)$ and the $\alpha$-regime corresponding to complete structural relaxation.}, we initially focus on the long time limit and work with $F(k)$ directly. We decompose the order parameter into two parts, $F(k) = F_\mathrm{MCT}(k) + \Delta F(k)$, where $F_{\mathrm{MCT}}(k)$ is the MCT solution. Then, we consider contributions to the correction term $\Delta F(k)$ originating from a class of diagrams that involve the three-point susceptibilities and are neglected in standard MCT. In this way we incorporate the influence of the fluctuations on the order parameter. An explicit perturbative analysis reveals that the resulting series grows increasingly singular near the transition at each order, signaling the complete breakdown of mean-field behavior below an upper critical dimension $d_c=8$. Notably, this result is fully consistent with replica field-theoretic approaches \cite{franz2011field}. 

In the second step, we resum this series by recasting it as an effective stochastic problem: we introduce a dynamical correlation function $F_u(\boldsymbol{k}_+,\boldsymbol{k}_-;t,0)$ evolving under spatio-temporally fluctuating external fields, and recover $F(k,t)$ through a disorder average of $F_u$. This systematically resums the dominant contributions to $\Delta F(k,t)$, formally extending MCT beyond its mean-field roots. 

Analyzing the time evolution of $F_u$ asymptotically in the $\beta$-regime yields an effective equation coinciding with the stochastic beta-relaxation theory \cite{rizzo2014long, rizzo2016dynamical, rizzo2020solvable}. Thus, our calculation provides, for the first time, a fully microscopic and dynamical derivation of this effective theory for a system of interacting particles, with all coupling constants determined from the static structure factor \emph{alone}. The final result is that the mean-field arrested phase is destabilized and replaced by a smooth crossover at all finite densities and temperatures\textemdash{}consistent with the absence of a true dynamical arrest observed in experiments and simulations, and without invoking activated processes or facilitation. 

\begin{figure*}
    \centering
    \includegraphics[width=0.85\textwidth]{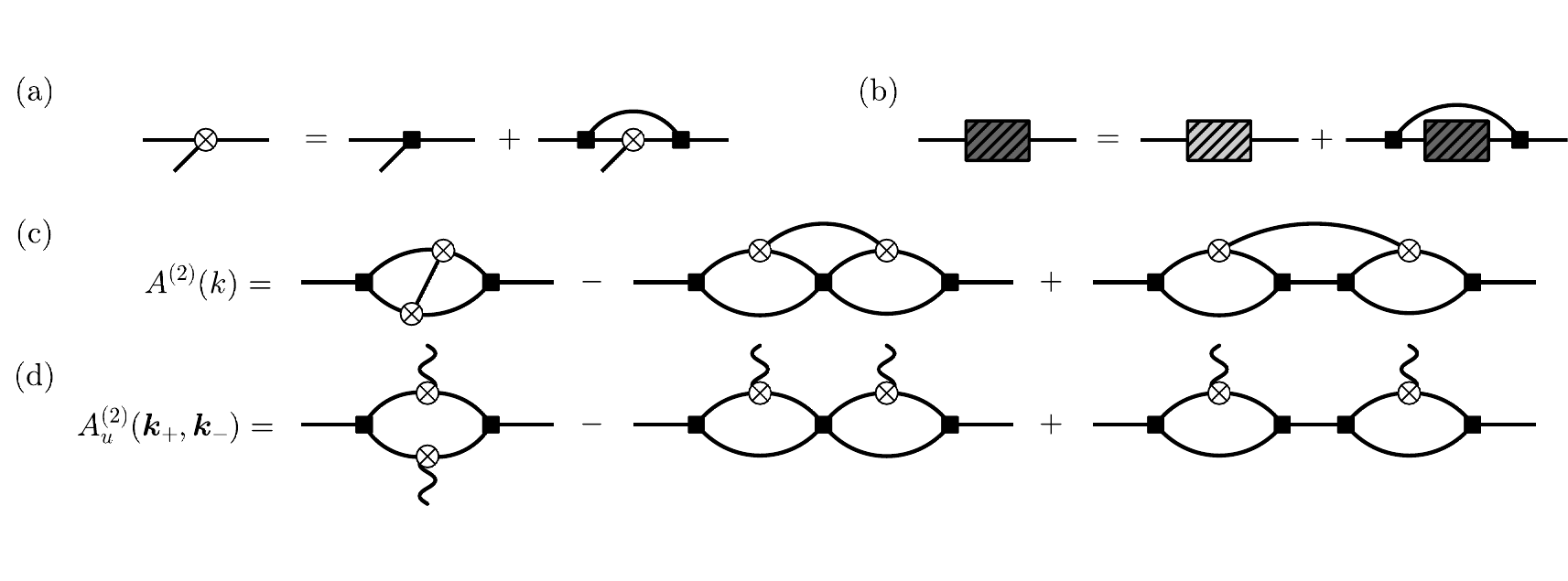}
    \vspace{-24pt}
    \caption{(a) Diagrammatic structure of Eq.~\eqref{eq:bethe_salpeter_3_vertex} for $\chi_{\boldsymbol{q}}^-(\boldsymbol{k})$, here represented by diagram\raisebox{-1.33ex}{\includegraphics[height=3ex,page=1]{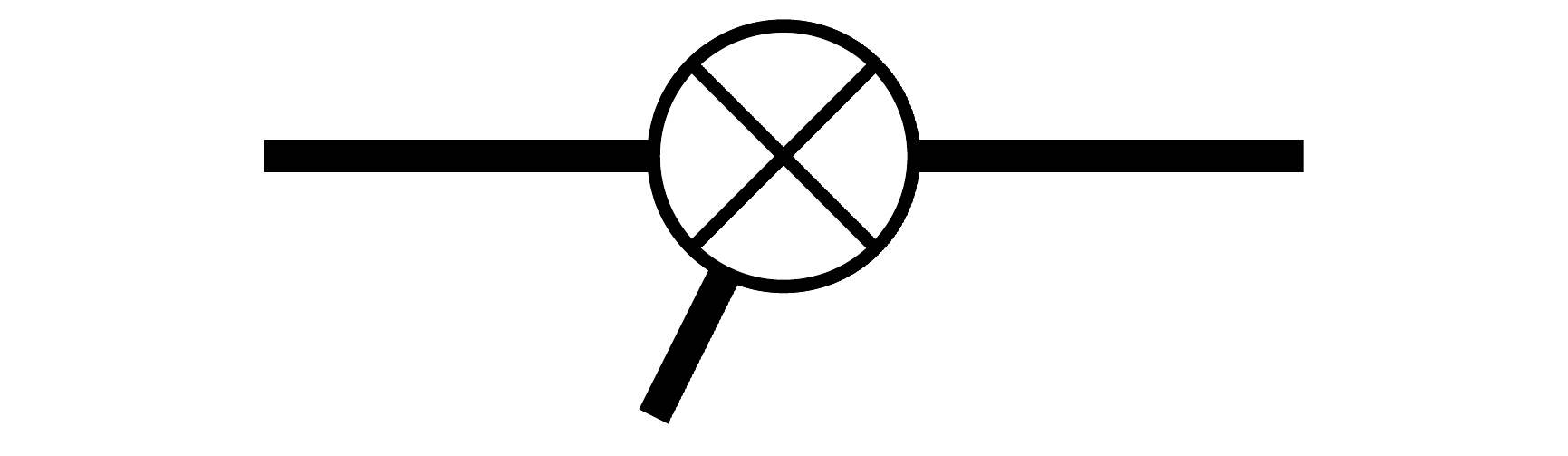}}. The $\chi_{\boldsymbol{q}}^+(\boldsymbol{k})$ susceptibility is obtained by using a right-handed vertex instead. Bonds represent the mode-coupling propagator $F_{\mathrm{MCT}}(k)$. (b) Diagrammatic structure of Eq.~\eqref{eq:generic_overarching}. The dark and light grey dashed box with two external legs represents $F^{(2n)}(k)$ and $A^{(2n)}(k)$ respectively. (c) All possible divergent diagrams with two connected rainbow insertions.
    (d) Second order effective sources $A_u^{(2)}(\boldsymbol{k}_+, \boldsymbol{k}_-)$ generated by the perturbative solution to Eqs.~\eqref{eq:postulate_microscopic_stochastic}–\eqref{eq:inhomogeneous_Mirr}. Note that \raisebox{-2.2ex}{\includegraphics[height=4ex,page=1]{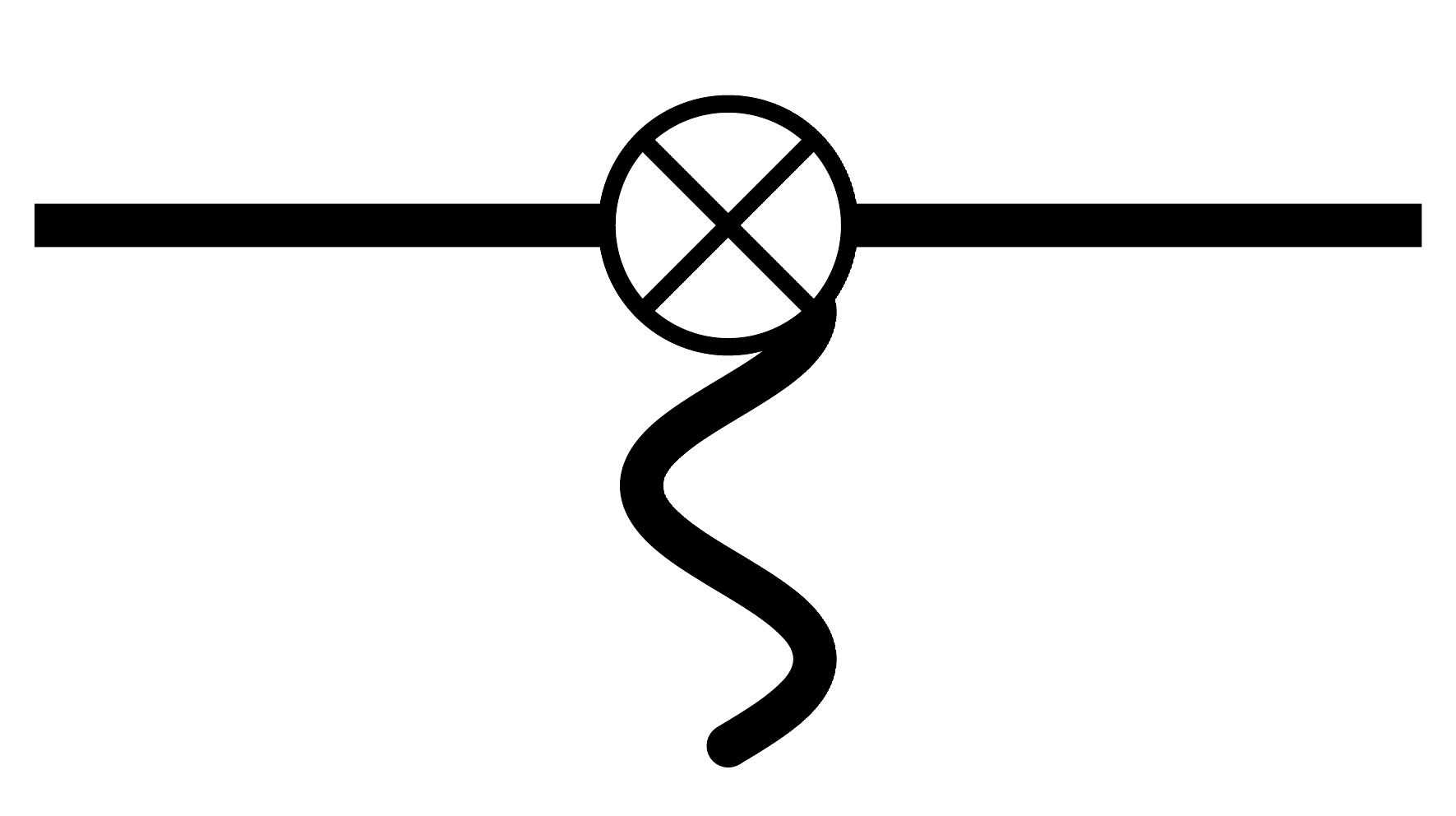}} $= \sum_{\nu=\pm}\chi_{\boldsymbol{q}}^{(\nu)}(\boldsymbol{k})u^{(\nu)}_{\boldsymbol{q}}$, where the\raisebox{-0.2ex}{\includegraphics[height=2ex,page=1]{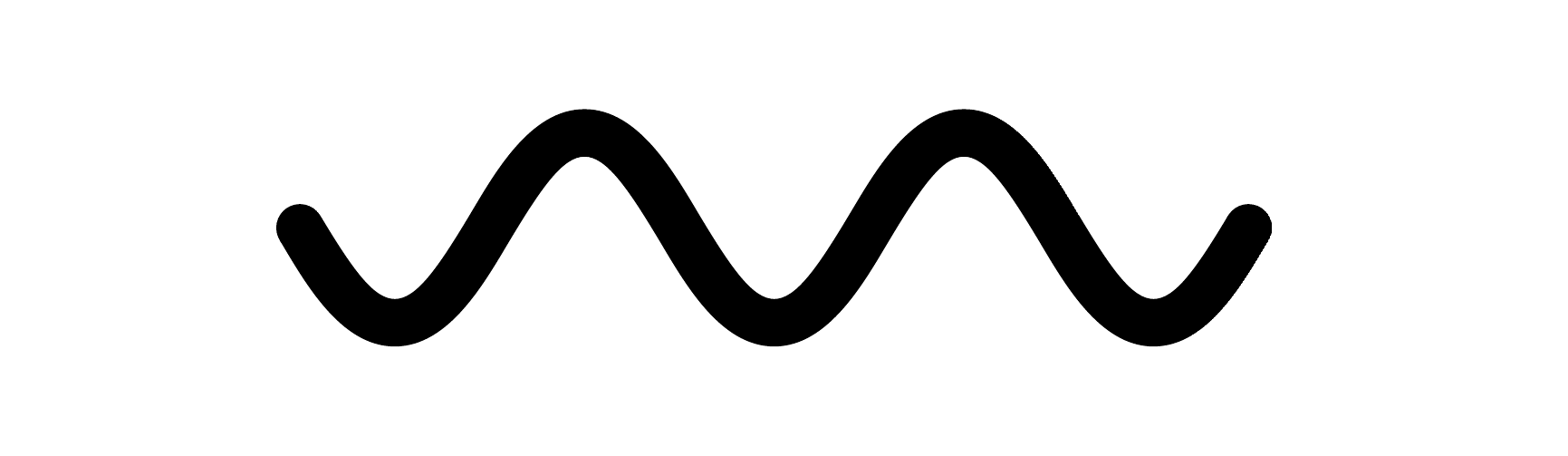}}line represents a sum over the random fields $u^{\pm}_{\boldsymbol{q}}$.
    }
\label{fig:main_fig}
\end{figure*}

\emph{Critical Fluctuations Around MCT}\textemdash{}In the present diagrammatic setting\footnote{Note that we use a set of slightly different but equivalent diagrammatic
rules compared to Ref.~\cite{szamel2007dynamics}.}, the key three-point susceptibility encoding the critical fluctuations  emerges naturally from the resummation of a specific class of diagrams referred to as `rainbow diagrams' \cite{szamel2008divergent, szamel2013breakdown}. In the long-time limit, two susceptibilities appear $\chi_{\boldsymbol{q}}^\pm(\boldsymbol{k})$, related by time-reversal symmetry \cite{szamel2008divergent, companionpaper}. The susceptibilities satisfy the linear integral equation
    \begin{equation}
        \chi_{\boldsymbol{q}}^\pm(\boldsymbol{k}) = \chi_{\boldsymbol{q}}^{(0\pm)}(\boldsymbol{k}) + \int \frac{\mathrm{d}\boldsymbol{p}}{(2\pi)^d}\, \mathcal{M}_{\boldsymbol{q}}(\boldsymbol{k}, \boldsymbol{p})\, \chi_{\boldsymbol{q}}^\pm(\boldsymbol{p})
        \label{eq:bethe_salpeter_3_vertex}
    \end{equation}
derived in Ref.~\cite{szamel2013breakdown} and shown diagrammatically in Fig.~\ref{fig:main_fig}(a). The source term $\chi_{\boldsymbol{q}}^{(0\pm)}(\boldsymbol{k})$ in Eq.~\eqref{eq:bethe_salpeter_3_vertex} can be expressed in terms of simple structural correlations such as the structure factor $S(k)$. 
The mass operator $\mathcal{M}_{\boldsymbol{q}}$ in Eq.~\eqref{eq:bethe_salpeter_3_vertex} was introduced by \citet{biroli2006inhomogeneous}, who showed that in the long-wavelength limit, $|\boldsymbol{q}|\to0$, it is proportional to MCT's so-called `stability operator', which controls the mode-coupling transition. In particular, a non-vanishing Debye-Waller factor $F(k)$ emerges when the largest eigenvalue of $\mathcal{M}_{\boldsymbol{q}\to\boldsymbol{0}}$ reaches unity. Let $n$ denote the number density, and $\varepsilon=(n-n_c)/n_c$  the relative distance from the transition. Since we work in the non-ergodic phase, $\varepsilon < 0$. In the limit of small $|\boldsymbol{q}|\equiv q$ and small distance $|\varepsilon|$, the solution to Eq.~\eqref{eq:bethe_salpeter_3_vertex} takes the form
    \begin{equation}
        \chi_{\boldsymbol{q}}^\pm(\boldsymbol{k}) = \frac{1}{|\varepsilon|^{1/2}} \frac{b_\pm S(k) h_0^{\mathrm{R}}(k)}{1 + (|\varepsilon|^{-1/4}\xi_0q)^2},
        \label{eq:X3_asymptotic_non_ergodic}
    \end{equation}
where $h_0^{\mathrm{R}}(k)$ is the right eigenfunction of the stability operator corresponding to the largest eigenvalue, 
calculated at the mode-coupling transition \cite{biroli2006inhomogeneous, szamel2013breakdown}. The constants $b_\pm$ and the bare lengthscale $\xi_0$ can be expressed in terms of known quantities. We read off from Eq.~\eqref{eq:X3_asymptotic_non_ergodic} the diverging correlation length $\xi = \xi_0 |\varepsilon|^{-1/4}$ identified by Biroli \textit{et al.} 
\cite{biroli2006inhomogeneous}.

\emph{Leading Order Divergences Around MCT}\textemdash{}We seek to find fluctuation-dominated corrections, $\Delta F(k)$, to the Debye-Waller factor predicted by MCT, around the MCT transition. First, we identify a diagrammatic series where each term consists of mode-coupling contributions \cite{szamel2007dynamics}, dressed with $n$ pairs of rainbow insertions, connected by $F_{\mathrm{MCT}}(q)$. As the resummation of rainbow insertions leads to susceptibilities $\chi_{\boldsymbol{q}}^\pm(\boldsymbol{k})$ that diverge near the transition, the contributions in which they appear become increasingly dominant. For instance, the three contributing diagrams with one pair of susceptibilities, shown in Fig.~\ref{fig:main_fig}(c), sum to a contribution whose leading contribution scales as $A^{(2)}(k) \sim a^{(2)}(k)S(k) |\varepsilon|^{(d - 4)/4}$ around the mode-coupling transition. Here, $a^{(2)}(k)$ is some regular amplitude and the superscript “(2)” indicates the number of rainbow insertions. Including $n$ such pairs gives a set of contributions with leading singular behavior $A^{(2n)}(k) \sim a^{(2n)}(k) S(k) |\varepsilon|^{n(d - 4)/4}$. 

Next, we note that terms $A^{(2n)}(k)$ can be further dressed with overarching rainbows, giving contributions $F^{(2n)}(k)$, related to $A^{(2n)}(k)$ by
    \begin{equation}
        F^{(2n)}(k) = A^{(2n)}(k) + \int \frac{\mathrm{d}\boldsymbol{p}}{(2\pi)^d} \mathcal{M}_{\boldsymbol{0}}(\boldsymbol{k}, \boldsymbol{p}) F^{(2n)}(p).
    \label{eq:generic_overarching}
    \end{equation}
The general diagrammatic structure of Eq.~\eqref{eq:generic_overarching} is illustrated in Fig.~\ref{fig:main_fig}(b). We recognize in this equation the special case $\boldsymbol{q}=\boldsymbol{0}$ of Eq.~\eqref{eq:bethe_salpeter_3_vertex}. The solution can therefore be written directly, producing perturbative corrections whose leading behavior is of of the form $F^{(2n)}(k) \sim  f^{(2n)}(k)S(k)|\varepsilon|^{n(d-4)/4 - 1/2}$ for some regular amplitudes $f^{(2n)}(k)$. Summing up the contributions, we get the asymptotic series
   \begin{equation}
    \begin{split}
        \Delta F(k) = \sum_{n=1}^{\infty} f^{(2n)}(k)S(k) |\varepsilon|^{n(d-4)/4 - 1/2}
    \end{split}
    \label{eq:asymptotic_series_alt}
    \end{equation}
for the dominant corrections to the mode-coupling predictions near the transition due to critical fluctuations. We stress that the amplitudes $f^{(2n)}(k)$ can be determined diagrammatically and are computable solely from microscopic structural inputs.

Below four dimensions ($d<4$), the expansion for the leading corrections Eq.~\eqref{eq:asymptotic_series_alt}, breaks down as each higher-order term grows more singular. To determine the upper critical dimension, we compare the fluctuation correction $\Delta F(k)$ with the change in the Debye-Waller factor predicted by MCT close to transition: $\delta F_{\mathrm{MCT}}(k) \propto \sqrt{|\varepsilon|}$ \cite{gotze1985properties}. Inspection of Eq.~\eqref{eq:asymptotic_series_alt} shows that critical fluctuations are dominant for all $d < 8$, identifying $d_c = 8$ as the upper critical dimension. This result is consistent with Ginzburg criterion analyses \cite{biroli2004diverging, biroli2007critical, franz2011field}.

Having identified a set of divergent corrections to MCT, the natural next step is to resum them. However, the absence of a characteristic scale beyond which terms in Eq.~\eqref{eq:asymptotic_series_alt} become subleading in physically relevant dimensions, combined with the lack of generic expressions for the amplitudes of the leading corrections, renders this a strenuous task.

\emph{Mapping to a Stochastic Equation}\textemdash{}To overcome these difficulties, we map the asymptotic series Eq.~\eqref{eq:asymptotic_series_alt} onto a stochastic process whose perturbative solution reproduces the dominant asymptotic contributions. This type of approach has been introduced in the analysis of disordered critical systems \cite{parisi1979random, parisi1981critical} and later successfully applied in replica-based approaches to the structural glass transition \cite{franz2011field, franz2012quantitative}. To carry through the mapping, it is convenient to consider the full dynamical problem. 

The key observation is that the dominant contributions arise from mode-coupling diagrams decorated with susceptibilities, with the latter corresponding to variations of the order parameter with respect to an external field. To systematically take into account these contributions, we introduce a scheme in which the order parameter evolves under random, spatiotemporally varying external fields $u^{\pm}_{\boldsymbol{p}}(t)$, where $u^+_{\boldsymbol{p}}(t)$ and $u^-_{\boldsymbol{p}}(t)$ respectively inject and remove momentum $\boldsymbol{p}$ at time $t$. The order parameter is then replaced by its inhomogeneous in time and space counterpart, $F_u(\boldsymbol{k}_+, \boldsymbol{k}_-; t, 0)$, with an off-diagonal basis $\boldsymbol{k}_\pm = \boldsymbol{k} \pm \boldsymbol{q}/2$ such that the net momentum added by the external fields is $\boldsymbol{q}$. The subscript ``$u$" emphasizes that the new observable depends on stochastic fields $u^{\pm}_{\boldsymbol{p}}(t)$.

To ensure that the stochastic process governing $F_u(\boldsymbol{k}_+, \boldsymbol{k}_-; t, 0)$ generates the leading terms, we argue (and show in the companion paper) that it must obey a non-linear integro-differential equation structurally similar to that of mode-coupling theory: 
\begin{widetext}
\begin{equation}
    \begin{split}
        &\frac{\partial }{\partial t}F_u(\boldsymbol{k}_+, \boldsymbol{k}_-; t, 0) + D_0 k_+ \int \frac{\mathrm{d}\boldsymbol{p}}{(2\pi)^d}\int_0^t \mathrm{d}\tau R_u(\boldsymbol{k}_+, \boldsymbol{p} ; t, \tau) \frac{p}{S(p)} F_u(\boldsymbol{p}, \boldsymbol{k}_-  ; \tau, 0) = \mathcal{S}^{+}_u(\boldsymbol{k}_+,\boldsymbol{k}_- ; t, 0) + \mathcal{S}^{-}_u(\boldsymbol{k}_+,\boldsymbol{k}_- ; t, 0)
    \end{split}
\label{eq:postulate_microscopic_stochastic}
    \end{equation}
where $D_0$ denotes the bare diffusion constant of a Brownian particle.\footnote{Although we focus on overdamped dissipative systems, the approach should apply to other dynamics as well.} The two source terms on the right-hand-side of Eq.~\eqref{eq:postulate_microscopic_stochastic} couple the random fields linearly to the order parameter $F_u(\boldsymbol{k}_+, \boldsymbol{k}_- ; t, 0)$. The time evolution is governed by a resolvent operator $R_u(\boldsymbol{k}, \boldsymbol{k}'; t, t') \equiv\ \left[(2\pi)^d\delta(\boldsymbol{k}-\boldsymbol{k}')\delta(t-t') + M_u^{\mathrm{irr.}}(\boldsymbol{k}, \boldsymbol{k}'; t, t') \right]^{-1}$ with a memory kernel
    \begin{equation}
    \begin{split}
        &M_u^{\mathrm{irr.}}(\boldsymbol{k}, \boldsymbol{k}' ; t, t')
        = \frac{nD_0}{2}\int \frac{\mathrm{d}\boldsymbol{p}}{(2\pi)^d}\frac{\mathrm{d}\boldsymbol{p}'}{(2\pi)^d} v_{\boldsymbol{k}}(\boldsymbol{p}, \boldsymbol{k}-\boldsymbol{p}) F_u(\boldsymbol{p}, \boldsymbol{p}' ; t, t')F_u(\boldsymbol{k}-\boldsymbol{p}, \boldsymbol{k}'-\boldsymbol{p}' ; t, t') v_{\boldsymbol{k}'}(\boldsymbol{p}', \boldsymbol{k}'-\boldsymbol{p}')
    \end{split}
    \label{eq:inhomogeneous_Mirr}
    \end{equation}
taken in a mode-coupling-like approximation \cite{szamel1991mode} where $n$ is the particle number density. The precise forms of the sources $\mathcal{S}_u^\pm$ and the vertices $v_{\boldsymbol{k}}(\boldsymbol{q}, \boldsymbol{q}')$ are given in the companion work \cite{companionpaper}.
\end{widetext}

The physical correlation function $F(k,t)$ is given by the averaged solution
    \begin{equation}
        F(k,t-t') = \llbracket F_u(\boldsymbol{k}_+, \boldsymbol{k}_- ; t, t' )\rrbracket
    \label{eq:relation_Fu_F}
    \end{equation}
where $\llbracket ... \rrbracket$ denotes an average over realizations of the external random fields. To reproduce the diagrammatic contributions, these random fields are taken to be colored Gaussian processes satisfying $\llbracket u_{\boldsymbol{q}}^{\pm}(t) \rrbracket = 0$ and $\llbracket u_{\boldsymbol{q}}^+(t) u_{\boldsymbol{q}'}^-(t') \rrbracket = F_{\mathrm{MCT}}(q, t-t')(2\pi)^d\delta(\boldsymbol{q}-\boldsymbol{q}')\Theta(t-t')$,
with $\Theta(t)$ the Heaviside function, while $\llbracket u_{\boldsymbol{q}}^+(t) u_{\boldsymbol{q}'}^+(t')\rrbracket = \llbracket u_{\boldsymbol{q}}^-(t) u_{\boldsymbol{q}'}^-(t')\rrbracket = 0$.  

The system defined by Eqs.~\eqref{eq:postulate_microscopic_stochastic}–\eqref{eq:inhomogeneous_Mirr} provides a self-consistent resummation scheme for the dominant corrections of interest. Specifically, a perturbative expansion in the fields $u_{\boldsymbol{q}}^{\pm}(t)$, after subsequent disorder averaging,  reproduces, order by order, the leading order divergences $\Delta F(k)$, Eq.~\eqref{eq:asymptotic_series_alt}, around the MCT solution. Expanding the long-time solution as
    \begin{equation}
        F_u(\boldsymbol{k}_+, \boldsymbol{k}_-) = F_{\mathrm{MCT}}(k)(2\pi)^d\delta(\boldsymbol{q}) + \sum_{n=1}^{\infty} F_u^{(n)}(\boldsymbol{k}_+, \boldsymbol{k}_-),
    \end{equation}
one finds that each $n$-th order term satisfies an equation of the form 
    \begin{equation}  
    \begin{split}      
    F_u^{(n)}(\boldsymbol{k}_+, \boldsymbol{k}_-) =&\ A_u^{(n)}(\boldsymbol{k}_+, \boldsymbol{k}_-) \\
    &\ + \int \frac{\mathrm{d}\boldsymbol{p}}{(2\pi)^d}\mathcal{M}_{\boldsymbol{q}}(\boldsymbol{k}, \boldsymbol{p})F_u^{(n)}(\boldsymbol{p}_+, \boldsymbol{p}_-) 
    \end{split}    
    \label{eq:generic_structure_perturbative_expansion}
    \end{equation}    
with $\boldsymbol{p}_\pm=\boldsymbol{p}\pm \boldsymbol{q}/2$ and 
where $A_u^{(n)}(\boldsymbol{k}_+, \boldsymbol{k}_-)$ are effective source terms depending explicitly on the random fields. The structure of Eq.~\eqref{eq:generic_structure_perturbative_expansion} is directly equivalent to Eq.~\eqref{eq:generic_overarching} upon averaging over the random fields. To fully verify the equivalence, we must examine the source contributions $A_u^{(n)}(\boldsymbol{k}_+, \boldsymbol{k}_-)$. The detailed calculation is presented in the compation paper \cite{companionpaper}; here we present just an illustrative example. At second order, there are three contributions to $A_u^{(2)}(\boldsymbol{k}_+, \boldsymbol{k}_-)$, shown diagrammatically in Fig.~\ref{fig:main_fig}(d). The structural similarity with the diagrams $A^{(2)}(k)$ identified earlier is evident. Taking the average over the random fields, we obtain that $\llbracket A_u^{(2)}(\boldsymbol{k}_+, \boldsymbol{k}_-)\rrbracket = A^{(2)}(k)$ and, consequently, that $\llbracket F^{(2)}_u(\boldsymbol{k}_+, \boldsymbol{k}_-)\rrbracket \sim f^{(2)}(k)|\varepsilon|^{-(d-4)/4-1/2}$ thus capturing the desired asymptotic structure to leading order near the mode-coupling transition. This procedure can be iterated at all orders, such that one obtains the equivalence
    \begin{equation}
        \llbracket \sum_{n=0}^{\infty} F_u^{(n)}(\boldsymbol{k}_+, \boldsymbol{k}_-) \rrbracket \sim F_{\mathrm{MCT}}(k) + \Delta F(k)
    \end{equation}
asymptotically close to the mean-field transition.

\emph{Generalized $\beta$-scaling}\textemdash{}We now turn to an asymptotic analysis of Eq.~\eqref{eq:postulate_microscopic_stochastic} in the $\beta$-regime at a distance $\varepsilon$ to the transition, where the resummed corrections dominate. To proceed, we introduce, in the spirit of the standard mode-coupling analysis, a generalized factorisation ansatz around the plateau of the fluctuating correlator, 
    \begin{equation}
        F_u(\boldsymbol{k}_+, \boldsymbol{k}_-, t) = F_c(k) (2\pi)^d\delta(\boldsymbol{q}) + H(k)g_u(q, t)
    \label{eq:stochastic_ansatz_general}
    \end{equation}
where $H(k)$ encodes the microscopic corrections to $F_c(k)$ inherited from the undelrying mode-coupling structure and $g_u(\boldsymbol{q}, t)$ are the noise-induced fluctuations around the plateau. Taking the mean-field (i.e., noise-free) result $H(k) = S(k)h_0^{\mathrm{R}}(k)$, substituting Eq.~\eqref{eq:stochastic_ansatz_general}, and performing a low-$q$ expansion in Eq.~\eqref{eq:postulate_microscopic_stochastic} eventually yields a closed dynamical equation for the deviations from the plateau that depends on the standard mode-coupling parameters $\lambda$ \cite{gotze1985properties} and $\Gamma$ \cite{biroli2006inhomogeneous}.\footnote{$\Gamma$ is related to the bare lengthscale $\xi_0$ from Eq.~\eqref{eq:X3_asymptotic_non_ergodic} through $\Gamma\propto\xi_0^2$.} The original random fields $u^\pm(\boldsymbol{x}, t)$ become then time-independent and contribute with weights that differ by a constant. This allows us to replace them by a single quenched field which after an additional re-scaling (to absorb the weights) becomes a quenched random field $s(\boldsymbol{x})$ satisfying $\llbracket s(\boldsymbol{x})\rrbracket  = 0$ and $\llbracket s(\boldsymbol{x}) s(\boldsymbol{x}')\rrbracket = \Delta\sigma^2\delta(\boldsymbol{x} - \boldsymbol{x}')$. In real space, the fluctuations (now denoted $g_s$ to emphasize dependence on the stochastic field $s(\boldsymbol{x})$) at position $\boldsymbol{x}$ obey
    \begin{equation}
    \begin{split}
         \sigma + s(\boldsymbol{x}) =&\ -\Gamma \nabla^2 g_s(\boldsymbol{x},t) - \lambda g_s(\boldsymbol{x},t)^2  \\
         &\ + \frac{\mathrm{d}}{\mathrm{d}t} \int_0^t \mathrm{d}\tau g_s(\boldsymbol{x},t-\tau) g_s(\boldsymbol{x},\tau),
    \end{split}
    \label{eq:kinetic_SBR}
    \end{equation}
where $\sigma \propto \varepsilon$ is known as the separation parameter. The squared-gradient term in Eq.~\eqref{eq:kinetic_SBR} naturally emerges from the spectrum of the operator $\mathcal{M}_{\boldsymbol{q}}(\boldsymbol{k}, \boldsymbol{p})$ in the asymptotic analysis \cite{biroli2006inhomogeneous}. Importantly, the coefficients $\Gamma, \lambda$ and the noise variance $\Delta\sigma^2$ can be calculated from the pair structure; explicit expressions are provided in the companion paper \cite{companionpaper}. We have computed these constants for the paradigmatic hard-sphere system in the Percus-Yevick approximation, and report $\lambda = 0.735$, $\Gamma = 0.0708$ and $\Delta\sigma^2 = 0.0045$. Remarkably, we recognize in Eq.~\eqref{eq:kinetic_SBR} the stochastic-beta relaxation (SBR) equation, a beyond-mean-field effective theory of glassy dynamics derived using either a combination of replica and supersymmetric
field-theoretic techniques \cite{rizzo2014long, rizzo2016dynamical} or from an expansion around a mean-field kinetically constrained model \cite{rizzo2020solvable}.

Physically, the correlation function $g_s(\boldsymbol{x},t)$ encodes mesoscopic spatio-temporal fluctuations of the van Hove function. Crucially, the presence of quenched disorder in Eq.~\eqref{eq:kinetic_SBR} destabilizes the mean-field ideal glass phase predicted by conventional MCT, whereby the long-time fluctuations $\lim_{t\to\infty}\llbracket g_s(\boldsymbol{x},t)\rrbracket$ diverge to $-\infty$, indicating a systematic departure from the plateau and the onset of structural relaxation \cite{rizzo2015qualitative}.

\emph{Discussion}\textemdash{}We have provided a microscopic characterization of the glass crossover in supercooled liquids. Our central finding is that the disappearance of the mean-field transition results solely from the inclusion of critical fluctuations: so-called activated events \cite{rusciano2025low} and dynamic facilitation \cite{scalliet2022thirty} are not needed to rationalize the absence of dynamical arrest. 

The resulting theory naturally captures the notion of dynamically self-induced disorder in structural glasses \cite{berthier2021self}, which we show to become effectively quenched in the $\beta$-regime. This provides a microscopic justification, \emph{a posteriori}, for the use of quenched disorder in simplified models of glassy dynamics such as $p$-spin systems \cite{kirkpatrick1987p, kirkpatrick1989scaling}. More broadly, the present work unifies the microscopic mode-coupling and replica field-theory approaches to the glass transition, providing a dynamical microscopic basis that was previously absent from landscape-based approaches. This offers a natural picture for the evolution of dynamic heterogeneities: they first appear as critical fluctuations upon approaching the avoided critical point from the mildly supercooled regime, and acquire the character of quenched disorder deeper into the supercooled regime.

We emphasize that the present scheme is controllable in the $\beta$-regime of relaxation asymptotically close to the avoided dynamical transition. Access to $\alpha$-regime predictions is only indirect, through MCT-like asymptotic analyses of the late $\beta$-regime. The relaxation at deep supercooling lies outside the scope of the present framework.

\emph{Acknowledgements}\textemdash{} CCLL and LMCJ acknowledge the Dutch Research Council (NWO) for financial support through a Vidi grant. GS acknowledges
the support of NSF Grant No. CHE 2154241. CCLL also warmly thanks CSU Chemistry for their hospitality during part of this work. We are indebted to Tommaso Rizzo for useful discussions and to Ilian Pihlajamaa for critical reading of the manuscript.

\bibliography{apssamp}
\end{document}